# Tunnel-rate controlled local heat distribution in mesoscopic circuits

[1]*Parvathy Gireesan,* [1]*Umang Soni,* [1]*Harikrishnan Sundaresan,* [2]*Giorgio Biasiol, and* [1]*Madhu Thalakulam**

[1] School of Physics, Indian Institute of Science Education and Research Thiruvananthapuram, Kerala 695551, India
[2] CNR-Consiglio Nazionale Delle Ricerche, Istituto Officina Dei Materiali, Area Science Park, Strada Statale 14 Km 163.5, Basovizza, Trieste, 34149, Italy

**ABSTRACT**

Solid-state quantum technologies, including qubits and quantum metrology circuits, demand milli-Kelvin operation to preserve fragile quantum states from classical noise. While the negligible electron-phonon coupling is the major impediment, reaching ~50 mK electron temperature is further suffered by the high electrical resistance and sub-micron-scale dimensions of typical devices, limiting conventional heat dissipation. Though the phonons are effectively frozen, thermoelectric techniques could offer a viable path for heat management. This work explores thermally driven electrical transport in a gated quantum dot (QD) on a GaAs/AlGaAs two-dimensional electron gas (2DEG), to control heat flow between the source and drain reservoirs. By exploiting the QD's discrete energy spectrum and tuneable tunnel rates, a precise control over the polarity and magnitude of the resulting thermoelectric current is demonstrated. A temperature difference of ~650 mK is maintained across the QD, a separation of ~400 nm, by tuning the tunnel-rates. An experimental gate pulsing method is also introduced to directly measure the electron temperature differences across the QD, bypassing the need for any theoretical fits. The results presented here show that tuneable tunnel barriers can be used for local heat control, and could lead to advanced quantum refrigerators that work efficiently in mesoscopic circuits.

*Corresponding author, madhu@iisertvm.ac.in

## 1. Introduction

Two-dimensional electron gas (2DEG)-based nanoscale devices in semiconductor heterostructures such as GaAs/AlGaAs and Si/SiGe have been a platform for hosting a variety of quantum circuits [1-10], owing to its superior electrical properties such as high carrier mobility, local control of transport properties by electrostatic gating, and scalability. Owing to the feeble energy-scales of most quantum phenomena and susceptibility to various classical noises present in the system, realization of quantum technologies on these systems requires very low electron-temperatures [1-10]. Even though these experiments are performed in sub-Kelvin cryogenic systems with extensive thermalization and noise filtering schemes [11-13], thermalization and cooling of the 2DEG down to a few mK at these temperatures is a major challenge. The weak acoustic-phonon scattering at lower temperatures (< 1 K) [14] and heat leaks due to electromagnetic noise keep the 2DEG electron temperature considerably above the lattice temperature. It has been reported that the cooling profile of locally heated electrons in a 2DEG extends over a few tens of micro-meters at lattice temperatures below 1 K [14]. Since the dimensions of these quantum devices are in submicron scales, it is advantageous to have on-chip mechanisms which allow local control over heat distribution within mesoscopic circuits in 2DEG. While 2DEG without any tunnel barriers can be cooled to low temperatures by extensive filtering schemes [11-13], removing heat from a high-impedance system consisting of multiple tunnel barriers is a non-trivial challenge. Since the electron-phonon coupling is negligible, thermo-electrically cooling and distributing heat in these systems would be a viable strategy.

Following its theoretical prediction [15], thermally driven transport on gated quantum dots (QDs) has been a topic of interest for the last few decades [16-22]. Later experiments in few-electron QDs at lower temperatures revealed that the discrete energy spectra of the QD can resolve the line shape of the difference in Fermi distributions at the reservoirs in first-order thermo-current [19, 23-25]. A growing body of experimental works justify the potential of QD-based thermal machines to be used as basic units for thermal management within the 2DEG. QD heat engines have been experimentally shown to achieve near-thermodynamic limits in efficiency [25], in agreement with theoretical predictions [23, 26]. Local on-chip refrigeration of Fermi reservoirs has been realised experimentally using energy-selective tunnelling in molecular junctions [27] and normal-metal – insulator – superconductor (NIS) tunnel junctions [28, 29, 30]. Recent experimental [31] and theoretical [32] works show how photon-assisted tunnelling through NIS junctions and normal-metal – QD – superconductor junctions can control the electron temperature of the normal-metal and the photon number in the superconducting resonator mode, highlighting the relevance of energy-selective tunnelling for heat distribution in QD-based circuits. The use of gated quantum dots as such for refrigeration is rarely addressed, except for a few theoretical studies [33-35], and an experimental study [36] where two QDs in series cool a 6 $\mu m^2$ reservoir in between by ~90 mK below the surrounding electron-temperature. The discrete QD energy spectrum and its tunnel coupling to electron reservoirs decide the magnitude and direction of tunnelling current through the system – both of which are electrically tuneable in gated semiconducting QDs. In addition to being scalable and easily integrable with mesoscopic circuits in 2DEG, the fact that the operation of QD thermal machines can be switched electrically between that of a heat engine and a refrigerator by the relative detuning between QDs brings up a new playing field in heat management and refrigeration in mesoscopic regime.

In this manuscript, we demonstrate the control of heat flow between two points in a GaAs/AlGaAs 2DEG by exploiting the discrete energy spectra and tunnel rates of a gated QD in between. We choose a QD and double tunnel barrier system as opposed to a quantum point contact (QPC), since a tunnel barrier does not offer energy filtering and a QPC does not offer independent control over its tunnel rate and energy level alignment. The QD is realized on the GaAs/AlGaAs heterostructure by electrostatic gating. We find that the difference in Fermi distributions and hence the electron temperatures between the source and drain regions drive a thermoelectric current, whose polarity depends on the alignment of the dot energy level with the chemical potential of the source and drain. By controlling the effective tunnel rate, we are able to maintain a temperature difference of ~ 650 mK between the source and drain regions. We also demonstrate a method to estimate the difference in Fermi distributions across the QD by gate pulsing, without depending on any theoretical fits to the thermoelectric current. Our observations demonstrate that one can exploit tunnel barriers and controlling the tunnel rates as a tool to control the heat distribution in mesoscopic circuits, and pave the way for realizing refrigeration in the quantum regime.

## 2. Results

Figure 1(a) shows the depletion-gate layout of the QD device, and the gates coloured in blue form the QD used in this experiment. The gate pairs CB-MB and CB-RB form the left and right tunnel barriers respectively, while RP is the plunger gate used for tuning the electrostatic potential of the dot. A bias-tee allows simultaneous application of DC and high-frequency voltages pulses to RP. The device is DC voltage-biased in two-probe configuration ($V_{SD}$), and the drain current $I_{SD}$ is recorded with a low-noise current pre-amplifier. All voltages applied refer to the drain contact. Transfer characteristics of both tunnel barriers exhibit a clear pinch-off of the channel. A representative pinch-off curve of the tunnel barriers is shown in Supporting Information S1(a).

Figure 1(b) shows a map of the source-drain current $I_{SD}$ in the $V_{SD}$ - $V_{RP}$ plane; the clear diamond-shaped regions of no current, the Coulomb diamonds, are the touch-stone signatures of the formation of a well-defined QD on our device [37-39]. From the dimensions of the Coulomb diamonds, we estimate charging energies $E_C$ ~ 1.23 meV and 1.85 meV respectively for the charge states labelled M and M-1 in Figure 1(b).

Figure 1(c) shows line-profiles of the Coulomb diamond plot along the gate-voltage axis, for various $V_{SD}$ values within a 200 µV voltage window around $V_{SD} = 0$. From Figures 1(b, c), we find that at low $V_{SD}$ values (within the two dashed-lines in Figure 1(b)), the coulomb blockade oscillations (CBO) exhibit an antisymmetric shape as opposed to the regularly observed symmetric periodic peaks [37-39]. The linecuts at $V_{SD} \sim 0$ (lower panel) and that at $V_{SD} \sim 100$ μV (upper panel) are plotted separately in Figure 1(d) for clarity; as we increase the $V_{SD}$ values in either direction, the antisymmetric feature gradually evolves into the regularly observed CBO peaks. The CBO is a manifestation of unidirectional resonant transmission from source to drain across the dot through successive charge states, as one sweeps the gate voltage. The antisymmetric peaks observed in Figures 1(c, d) represents a switch in the transmission direction through each dot level, as the levels are swept across the Fermi-level.

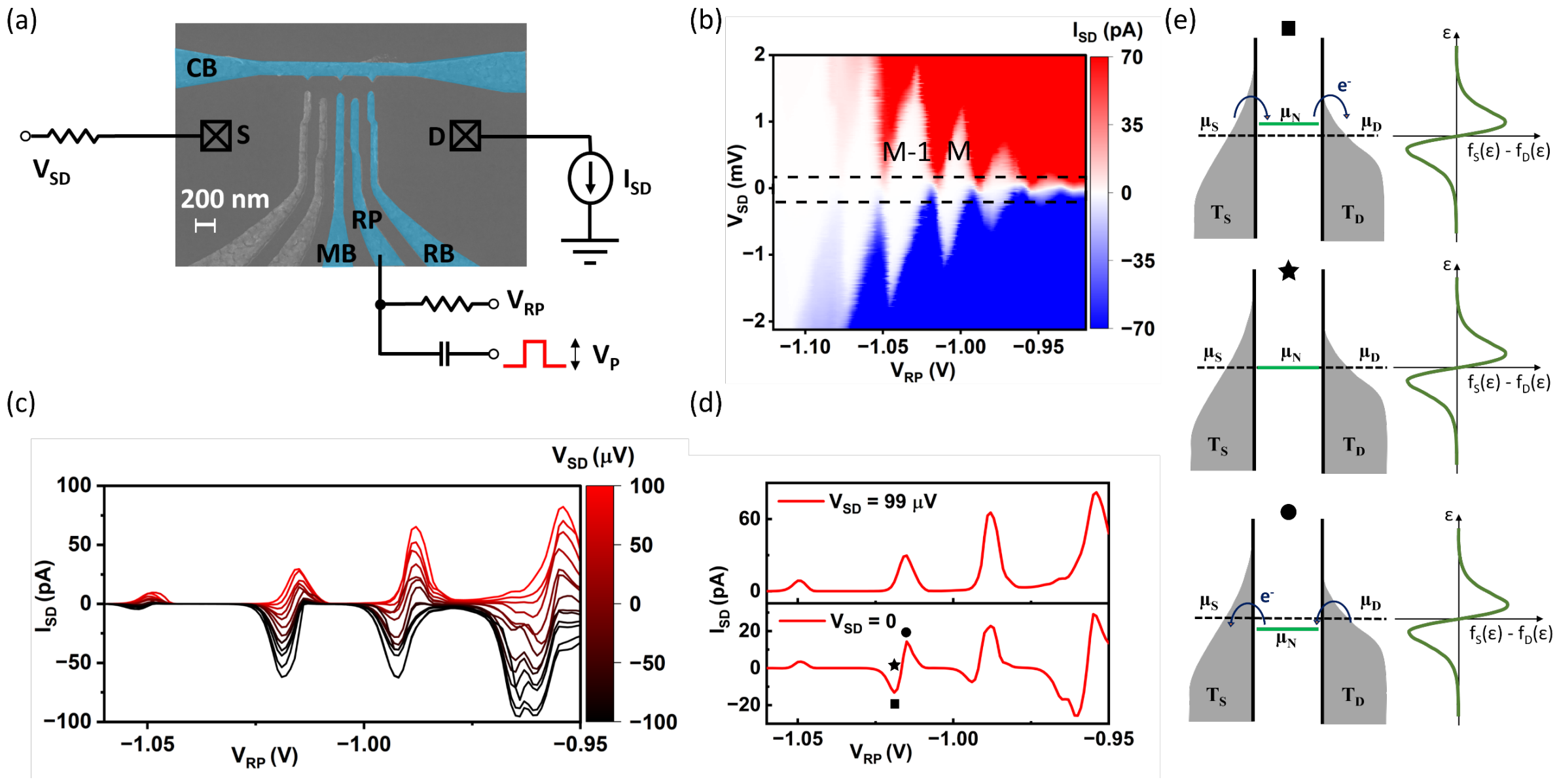

FIG.1. Observation of thermoelectric current in the device: (a) SEM image of the gate structure, with the circuit used in the experiment. Source and drain ohmic contacts are marked S and D respectively. (b) $I_{SD}$ as a function of $V_{SD}$ and $V_{RP}$, showing the Coulomb diamond feature. (c) Evolution of the CBO feature within a 200 μV window around $V_{SD} = 0$, taken from linecuts to (b). The $V_{SD}$ values corresponding to the $I_{SD}$-$V_{RP}$ sweeps are colour coded. (d) Two $I_{SD}$-$V_{RP}$ sweeps taken from (c) for clarity, show the bias-driven transport (positive source-drain bias, top panel) and thermally driven transport (zero source-drain bias, bottom panel). (e) Energy level diagrams for a QD with $T_S > T_D$ at $V_{SD} = 0$ corresponding to the three gate voltage values marked in (d), explaining the observed thermally driven transport. $f_S(\varepsilon) - f_D(\varepsilon)$ is plotted in green along the QD electrochemical potential ε.

Directional transport across the QD requires an asymmetry in the system, which is usually provided by the source-drain bias. Another source of asymmetry can arise from unequal tunnel barriers forming the dot, which has been shown to cause noise-rectification in the system [16, 40-43] – in this case the observed polarity of the conduction feature at $V_{SD} = 0$ should reverse when source and drain are interchanged. However, in our case the conduction feature retains its polarity (Supporting Information S1(b)), ruling out this possibility.

For symmetric tunnel barriers, a net tunnelling current requires a difference between the Fermi distributions at source and drain – realised experimentally by a voltage bias or a difference in electron temperatures between source and drain. The source-drain current dominated by first-order tunnelling through a single level QD as a function of the QD electrochemical potential ε is given by [23]

$$I_{SD}(\varepsilon) = -e \cdot \frac{\Gamma_L \Gamma_R}{\Gamma_L + \Gamma_R} \cdot [f_S(\varepsilon) - f_D(\varepsilon)] \quad (1)$$

where $\Gamma_L$ ($\Gamma_R$) is the tunnel coupling of the energy level to source (drain), $f_{S/D}(\varepsilon) = \left(1 + exp\left[\frac{\varepsilon - \mu_{S/D}}{K_B T_{S/D}}\right]\right)^{-1}$ is the Fermi distribution at source/drain, $T_S$ ($T_D$) and $\mu_S$ ($\mu_D$) are the electron temperature and Fermi level of source (drain). $\Gamma_{eff} = \frac{\Gamma_L \Gamma_R}{\Gamma_L + \Gamma_R}$ gives the effective tunnel rate across the QD. In the absence of any applied source-drain voltage bias, Equation-1 gives the thermoelectric (TE) current through the QD.

Here, we show that the polarity of the observed structure of CBO oscillations at $V_{SD} = 0$ in Figures 1(c, d) can be explained as the TE-current across the QD when $T_S > T_D$. Figure 1(e) shows the energy level diagrams corresponding to three different electrochemical potentials of the QD across the CBO resonance (accessed by varying $V_{RP}$, as marked in the lower panel of Figure 1(d)) at $V_{SD} = 0$. Though the source-drain bias is zero, the unequal electron temperatures at the source and drain contacts define a transport window. From Equation-1, the magnitude and direction of this current is decided by the difference in the Fermi functions, $f_{S/D}(\varepsilon)$ (plotted in green along the QD electrochemical potential in Figure 1(e)). As illustrated in Figure 1(e), the resulting current will be from drain to source when the dot level is above the Fermi-level (represented by the solid square), zero when the level is aligned with the Fermi-level (represented by solid star), and from source to drain when the level is below the Fermi-level (represented by the solid circle). The fact that the TE-feature retains its polarity upon interchanging source and drain (Supporting Information S1(b)) now implies that the source-side always stays hotter, suggesting that the noise reaching the sample from the measurement circuit is heating the source-side.

$T_S$ and $T_D$ can be obtained by fitting the TE-current at zero bias ($\mu_S = \mu_D = 0$) with Equation-1, once $\Gamma_L$ and $\Gamma_R$ are known. We use the saturation value of bias-driven current at $|V_{SD}| > K_B T/e$, assuming sequential first-order tunnelling through a single-level, to calculate $\Gamma_L$ and $\Gamma_R$ [44]. These saturation currents ($I_{+/-}$) correspond to the first step in the $I_{SD}$-$V_{SD}$ curve of the QD after Coulomb blockade is lifted in each biasing direction, and are given by

$$I_+ = 2e \frac{\Gamma_L \Gamma_R}{2\Gamma_R + \Gamma_L} \quad (2)$$

$$I_- = -2e \frac{\Gamma_L \Gamma_R}{\Gamma_R + 2\Gamma_L} \quad (3)$$

Equations 1, 2 and 3 characterise the TE-current and give $\Gamma_L$, $\Gamma_R$, $T_S$, and $T_D$, as demonstrated in Figure 2 for the charge transition shown in Figure 2(a). Figures 2(b, c) show the $I_{SD}$-$V_{SD}$ curves for positive and negative $V_{SD}$ respectively, obtained from linecuts to the Coulomb diamond at the $V_{RP}$ values marked by black dashed lines in Figure 2(a). The first plateau in the $I_{SD}$-$V_{SD}$ curve (marked in both figures) give the saturation currents to be $I_+ = 578$ pA and $I_- = -610$ pA, which are used to calculate $\Gamma_{eff} = 2.807$ GHz from Equations 2 and 3. The TE-feature given by the linecut along ε (marked by the green arrow in Figure 2(a)) is plotted in blue in Figure 2(d). Here, the electrochemical potential $\varepsilon_{RP}$ is calculated from $V_{RP}$ using the lever arm of ~0.0445, obtained from the Coulomb diamond; $\varepsilon_{RP} = 0$ corresponds to the resonance point of the charge transition. The red solid line in Figure 2(d) shows the fit to the data with Equation-1 using $\Gamma_{eff} = 2.807$ GHz, giving $T_S = 330$ mK and $T_D = 318$ mK.

Since TE-current is a strong function of tunnel rates, we now explore the possibility of using the QD's discrete energy spectrum and tunnel rates for managing heat transfer and distribution within the 2DEG. We use white noise to generate heat at the source-side, and observe the electron-temperature at the drain-side as a function of the effective tunnel rate $\Gamma_{eff}$. A bias-tee at room temperature is used to combine the DC bias $V_{SD}$ with the white noise generated by an arbitrary wave form generator, with a bandwidth of 80 MHz. $\Gamma_{eff}$ is varied from 575 MHz to 2.11 GHz by tuning the voltages on the gates CB, MB, RB, and RP. Figure 3 summarizes the effect of $\Gamma_{eff}$ on $T_S$, $T_D$, and $\Delta T = T_S - T_D$, for different heating powers on the source-side.

Figure 3(a) shows the variation in $\Delta T$ with $\Gamma_{eff}$, for different white-noise amplitudes, $V_{WN}$. We find that the temperature difference across the QD can be controlled by the effective tunnel rate; $\Delta T$ clearly increases as $\Gamma_{eff}$ is decreased at constant heating power. The inset to Figure 3(a) shows the fractional increase in $\Delta T$ ($[\Delta T_{0.575\text{ GHz}} - \Delta T_{2.11\text{ GHz}}] / \Delta T_{2.11\text{ GHz}}$) for $\Gamma_{eff}$ between 0.575 GHz and 2.11 GHz, against the heating amplitude $V_{WN}$. We see that $\Delta T$ shows a fractional increase of ~ 4.92 at $V_{WN} = 40$ $mV_{PP}$, but shows no measurable change at $V_{WN} = 0$, suggesting that the extent of this control over the heat flow decreases as the temperature difference $\Delta T$ decreases. When $\Delta T$ becomes comparable to the lattice temperature of 10 – 15 mK, $\Gamma_{eff}$ seems to have minimal effect on controlling the heat flow.

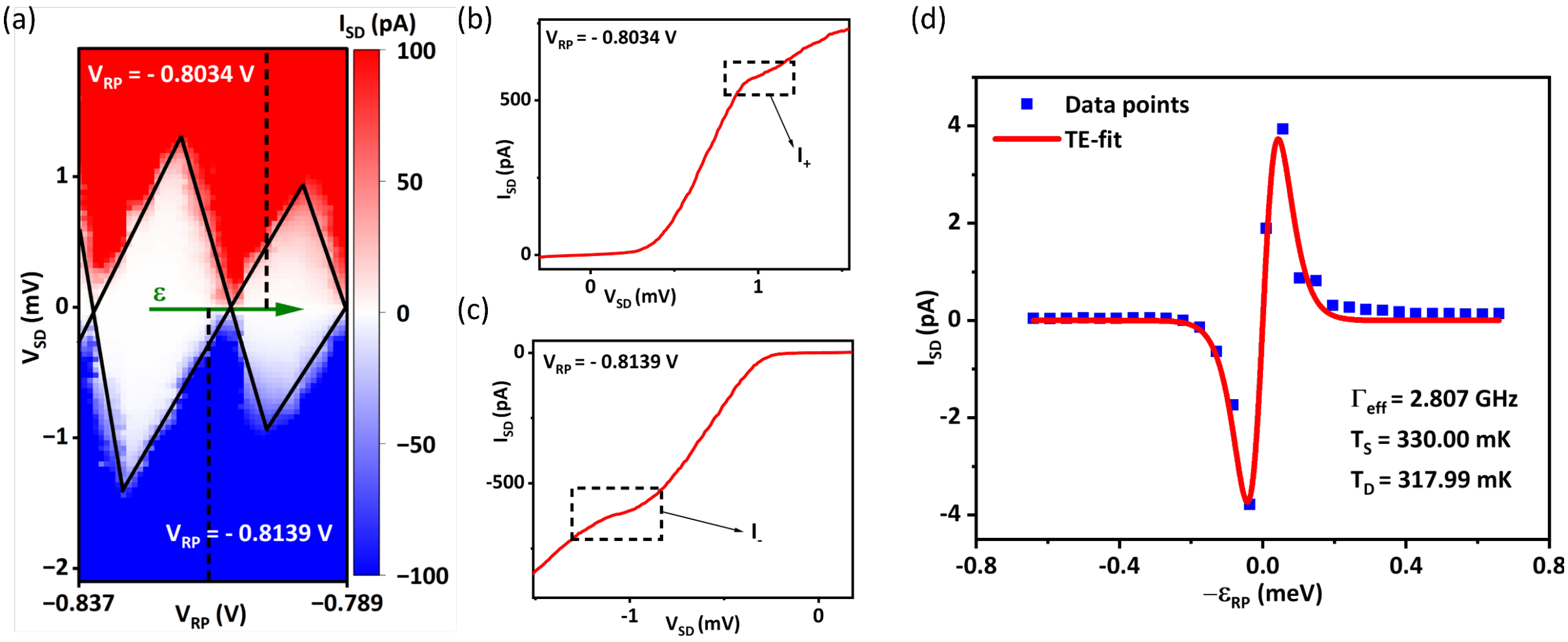

FIG. 2: Characterising the TE-current: (a) The CBO feature in between the two Coulomb diamonds is the regime of interest. Black solid lines outline the two diamonds. (b, c) $I_{SD}$-$V_{SD}$ curves given by linecuts at the two $V_{RP}$ values marked by black dashed lines in (a). The first plateau in $I_{SD}$ give the saturation currents $I_{+/-}$ as marked. (d) The TE-feature observed in this regime, given by the linecut along ε (green arrow in (a)) is plotted in blue and fitted with Equation 1.

Figures 3(b) and 3(c) show the responses of $T_S$, $T_D$, and $\Delta T$ to $\Gamma_{eff}$ at $V_{WN}$ = 20 and 40 $mV_{PP}$ respectively. We find that increasing $\Gamma_{eff}$ enhances the thermalisation between source and drain and causes an overall decrease in $T_S$ and $\Delta T$ in accord. However, $T_D$ is seen to initially shoot up as $\Gamma_{eff}$ is increased from 575 MHz to 735 MHz (regions shaded in yellow in Figures 3(b, c)), and decrease afterwards. This implies that the heat transfer for $V_{WN} \geq 20$ $mV_{PP}$ is limited by tunnel rate at $\Gamma_{eff} = 0.575$ GHz, but not at $\Gamma_{eff} \geq 0.735$ GHz. The $\Gamma_{eff}$-dependence at $V_{WN}$ = 40 $mV_{PP}$ shows the same trend as $V_{WN}$ = 20 $mV_{PP}$, but with a higher initial jump in $T_D$ – implying that the heating power due to $V_{WN}$ = 40 $mV_{PP}$ is still lower than the maximum heat that can be transferred at $\Gamma_{eff}$ = 0.735 GHz. Once the heat transfer is not limited by tunnel rate (at $\Gamma_{eff} \geq 0.735$ GHz and $V_{WN} \leq 40$ $mV_{PP}$), increasing $\Gamma_{eff}$ decreases both $T_S$ and $T_D$, as the two reservoirs progressively thermalise better with each other and get cooled by the cryostat's cooling power acting on both sides. Hence, for the heating powers used in this study, $\Gamma_{eff}$ = 0.575 GHz gives a strong control over heat transfer across the QD, while this control is lost for $\Gamma_{eff} \geq$ 735 MHz.

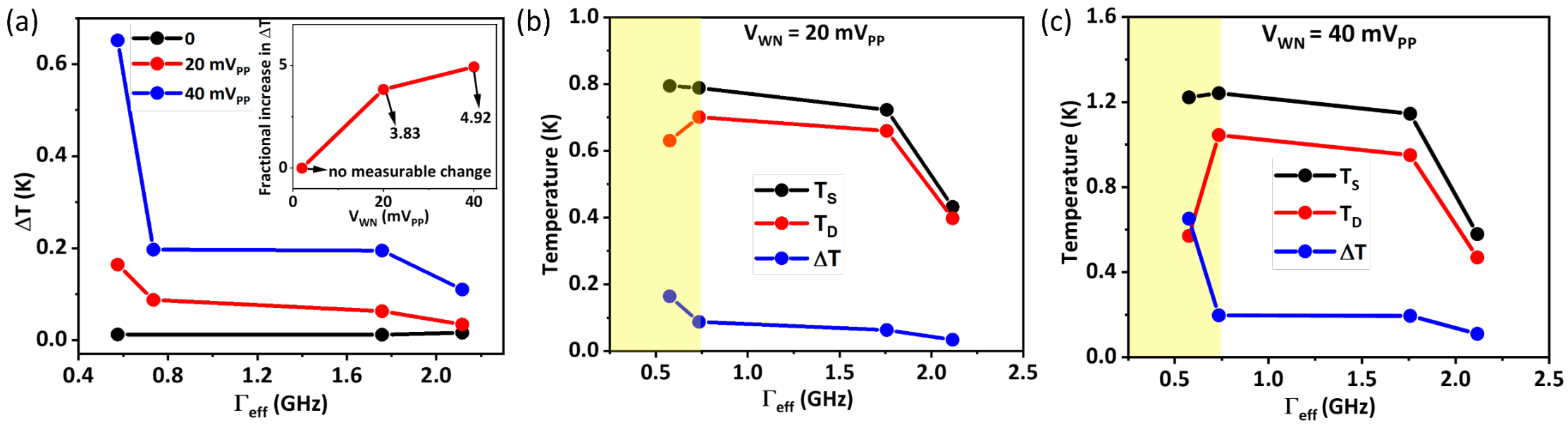

FIG. 3: $\Gamma_{eff}$ controls the heat flow to drain-side: (a) $\Delta T$ as a function of $\Gamma_{eff}$, for three different $V_{WN}$ values. Inset: The fractional increase in $\Delta T$ from $\Gamma_{eff}$ of 2.11 GHz to 0.575 GHz is plotted as a function of $V_{WN}$. The dependence of $T_S$, $T_D$, and $\Delta T$ on $\Gamma_{eff}$ is shown for $V_{WN}$ of (b) 20 $mV_{PP}$ and (c) 40 $mV_{PP}$. The $V_{WN}$ - $\Gamma_{eff}$ regimes where $\Gamma_{eff}$ controls heat flow are shaded in yellow. The solid lines in (a-c) are a guide to the eye.

In Figure 4, we exploit the gate pulsing technique to get an independent estimate of $f_S(\varepsilon) - f_D(\varepsilon)$ without relying on tunnel rate calculations. We apply a 1 MHz square-pulse of amplitude $V_P$ along with the D.C. gate voltage $V_{RP}$ on to the plunger gate RP, using a bias-tee mounted on the mixing chamber of the cryostat. Figure 4(a) shows the dot current $I_{SD}$ recorded as a function of $\varepsilon_{RP}$ and $V_P$ in the absence of any white-noise, at $V_{SD}$ = -17 μV and $V_{WN}$ = 0, where the transport is bias-driven. All the charge transitions split into two arms, the upper and lower arms correspond to the low and high states of the pulse respectively [45, 46].

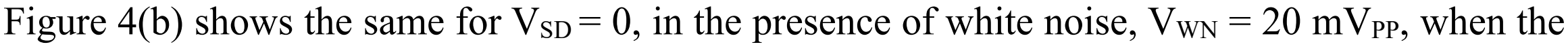

Figure 4(b) shows the same for $V_{SD} = 0$, in the presence of white noise, $V_{WN}$ = 20 $mV_{PP}$, when the transport is thermally driven. We focus on the charge transition enclosed by the green dashed lines in Figure 4(b), which corresponds to $\Gamma_{eff}$ = 2.11 GHz. The lever arm from the corresponding Coulomb diamond calibrates $V_{RP}$ into energy $\varepsilon_{RP}$ , while the peak splitting with $V_P$ in bias-driven transport (marked by black solid lines in Figure 4(a)) calibrates $V_P$ into energy $\varepsilon_P$ (calibration details are given in Supporting Information S2(a)). TE-current characterisation at $\varepsilon_P = 0$ from Figure 4(b) gives $T_S$ and $T_D$ to be 432 mK and 398 mK respectively (details of the TE-fit are given in Supporting Information S2(b)).

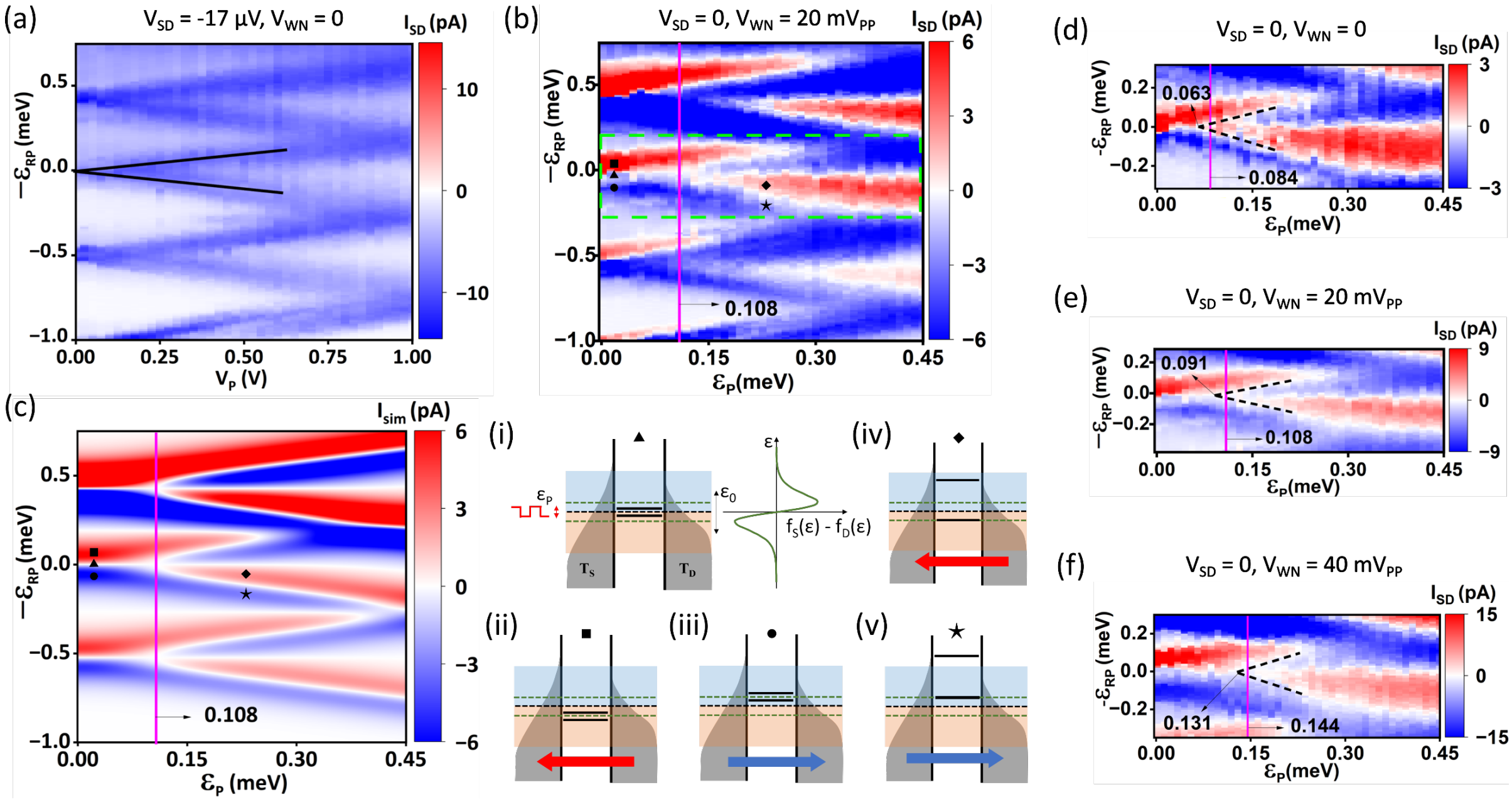

FIG. 4: Thermoelectric effect as seen in pulsed gate spectroscopy: Pulsed gate spectroscopy of charge states in the same gate-voltage regime, when transport is (a) bias-driven and (b) thermally driven. (c) shows the simulation of the data in (b). The branching pattern observed in (b) and (c) are explained by energy level diagrams (i-v) corresponding to the marked points. The direction of electron tunneling which gives positive (negative) current is marked in solid red (blue) arrows. Thermally driven pulsed-gate spectroscopy data for the transition at $\varepsilon_{RP} = 0$ for $V_{WN}$ of (d) 0, (e) 20 $mV_{PP}$, and (f) 40 $mV_{PP}$, show the calculated (magenta solid line) and measured (intersection of black dashed lines) values of the branching point shifting to higher $\varepsilon_P$ values as $V_{WN}$ is increased.

We simulate the experimental data in Figure 4(b) by taking into account the TE-current (from Equation-1) due to the three successive charge transitions at $\varepsilon_{RP}$ = 0.4279, 0, and -0.5238 meV, and the obtained TE-current $I_{sim}$ (see Supporting Information S3, Equation S3-2) is plotted in Figure 4(c). We observe branching in the thermally (noise) driven pulsed-gate spectroscopy data (Figure 4(b)) and its corresponding simulation (Figure 4(c)). We explain these transport features with the help of the QD energy level diagrams shown in Figure 4, Illustrations (i-v). Illustration-(i), represented by the filled triangle shows the level diagram and the splitting of the energy level under very low pulse amplitudes. The TE-current depends on the difference $f_S(\varepsilon) - f_D(\varepsilon)$, which is represented by the green line in Illustration-(i). $\varepsilon_0$ is the separation between the two extrema of $f_S(\varepsilon) - f_D(\varepsilon)$, marked by dotted green lines in illustrations (i-v). The two black lines mark the positions of the dot energy level during the high and low stages of the voltage pulse. The value of $\varepsilon_0$ calculated from the TE-fit (see Supporting Information S4(b)) is marked by magenta vertical line in Figures 4(b, c).

For $\varepsilon_{RP} = 0$, the dot level is resonant with the Fermi levels, and the two displaced positions of the dot level are placed symmetrically with respect to $\mu_N$, giving zero net current irrespective of the value of $\varepsilon_P$, as shown in Illustration-(i). The peaks in conduction are observed when the two displaced positions of the dot level align with the extrema of $f_S(\varepsilon) - f_D(\varepsilon)$, during gate-pulsing. When $\varepsilon_P < \varepsilon_0$ (to the left of the magenta vertical line in Figures 4(b, c)), the level splitting is not large enough for their thermally-driven conduction peaks to be resolved separately. This gives conduction peaks of opposite polarity above and below $\varepsilon_{RP} = 0$, as shown in Illustrations-(ii and iii). The red/blue solid arrows in the energy level diagrams indicate the direction of net electron tunnelling in each case.

When $\varepsilon_P > \varepsilon_0$, sweeping $\varepsilon_{RP}$ allows the extrema of $f_S(\varepsilon) - f_D(\varepsilon)$ to be resolved separately for the two positions of the energy level – giving two sets of such conduction peaks as shown in Figures 4(b) and (c), whose separation scales linearly with $\varepsilon_P$. TE-current is observed in the forward (red arrow, represented by solid diamond) and backward (blue arrow, represented by star) directions when either of the two displaced positions of the dot level align with the lower and upper extrema of $f_S(\varepsilon) - f_D(\varepsilon)$, as shown in Illustrations-(iv) and (v) respectively. The branching observed experimentally in Figure 4(b) is seen to quantitatively match its simulation in Figure 4(c), within the experimental uncertainties.

We note that the value of $\varepsilon_0$, which is a measure of $f_S(\varepsilon) - f_D(\varepsilon)$, can be obtained directly from this distinct branching pattern in the thermally-driven pulsed gate spectroscopy data, without depending on any theoretical fits to the TE-current. Figures 4(d, e, f) compare the calculated and measured values of $\varepsilon_0$, for different heating amplitudes ($V_{WN}$) at the source-side. The $\varepsilon_0$ values calculated from the respective TE-fits (see Supporting Information S4(a, b, c)) are marked by magenta solid lines in Figures 4(d, e, f). A rough estimate of $\varepsilon_0$ is obtained from the intersection of the zero-current lines in the two halves of the branching pattern (marked by black dashed lines in Figures 4(d, e, f)). Figures 4(d, e, f) show both the measured and calculated values of $\varepsilon_0$ shifting to higher $\varepsilon_P$ as $V_{WN}$ is increased, reflecting the increased thermal broadening. We find that the measured value of $\varepsilon_0$ falls within 21 μeV of the calculated value at all heating amplitudes, agreeing well given the experimental uncertainties.

## 3. Conclusions

In this manuscript, we study the tunnel rate assisted local heat control across a gated QD on GaAs/AlGaAs. For different source and drain temperatures, the tunnel barriers forming the QD and the discrete energy spectrum serve as an energy filter to drive a current through the system, under no applied bias. The effective tunnel rate is seen to strongly control the temperature difference $\Delta T$ across the QD, when $\Delta T$ is much higher than the lattice temperature. We inspect the tunnel rate assisted thermal isolation across the device, by applying white-noise to preferentially heat the source electrode and inspecting the drain temperature as a function of the tunnel rate. We find that for lower tunnel rates, ~575 MHz, the heat applied at the source-side does not get transferred to the drain-side beyond a threshold value, while for higher tunnel rates in the range of ~ 735 MHz to 2.11 GHz, the thermalization between the source and the drain sides improves. Additionally, we present a strategy to map the difference in the Fermi-Dirac distributions between the source and the drain sides, using gate pulsing, eliminating the need for theoretical fitting of thermoelectric current. The results presented in this manuscript emphasize the promise of adjustable tunnel barriers as a key technique for local heat management, and suggest new possibilities for highly efficient quantum refrigerators in mesoscopic circuits.

## 4. Experimental methods

The device is fabricated on a doped GaAs/AlGaAs heterostructure, hosting a 2DEG ~ 90 nm below the wafer surface with a carrier density and mobility of ~ $1.75 \times 10^{11}$ cm$^{-2}$ and ~ $1.5 \times 10^{6}$ cm$^{2}$/Vs respectively at 4 K. The wafer is wet-etched with citric acid/$H_2O_2$ to define the device mesa. Indium hot joints made to the wafer are annealed at 400 $^{0}$C in forming-gas atmosphere to achieve ohmic contacts to the 2DEG. The QD is fabricated by split-gate technique using photolithography and electron beam lithography, followed by Cr/Au metallization. The sample is mounted at the mixing chamber plate of a cryogen-free dilution refrigerator. All measurements are done at a base temperature of ~ 10 mK.

**Acknowledgements:** MT acknowledges receiving funding support from the National Quantum Mission, an initiative of the Department of Science and Technology, Ministry of Science and Technology, India, under the Grant No. DST/QTC/NQM/QC/2024.

**Author contributions:** MT conceived the problem, GB grew the GaAs/AlGaAs heterostructures, PG fabricated the devices with the help of HS, and did the measurements with the help of US. PG did the data analysis, simulations and wrote the initial draft of the manuscript, and PG and MT co-wrote the final draft.

**Conflicts of Interest:** The authors declare no conflict of interest.

**Data Availability Statement:** The data that support the findings of this study are available from the corresponding author upon reasonable request.

**Supporting Information**

**Tunnel-rate controlled local heat distribution in mesoscopic circuits**

[1]*Parvathy Gireesan,* [1]*Umang Soni,* [1]*Harikrishnan Sundaresan,* [2]*Giorgio Biasiol, and* [1]*Madhu Thalakulam*[*]

[1]School of Physics, Indian Institute of Science Education and Research Thiruvananthapuram, Kerala 695551, India
[2] CNR-Consiglio Nazionale Delle Ricerche, Istituto Officina Dei Materiali, Area Science Park, Strada Statale 14 Km 163.5, Basovizza, Trieste, 34149, Italy

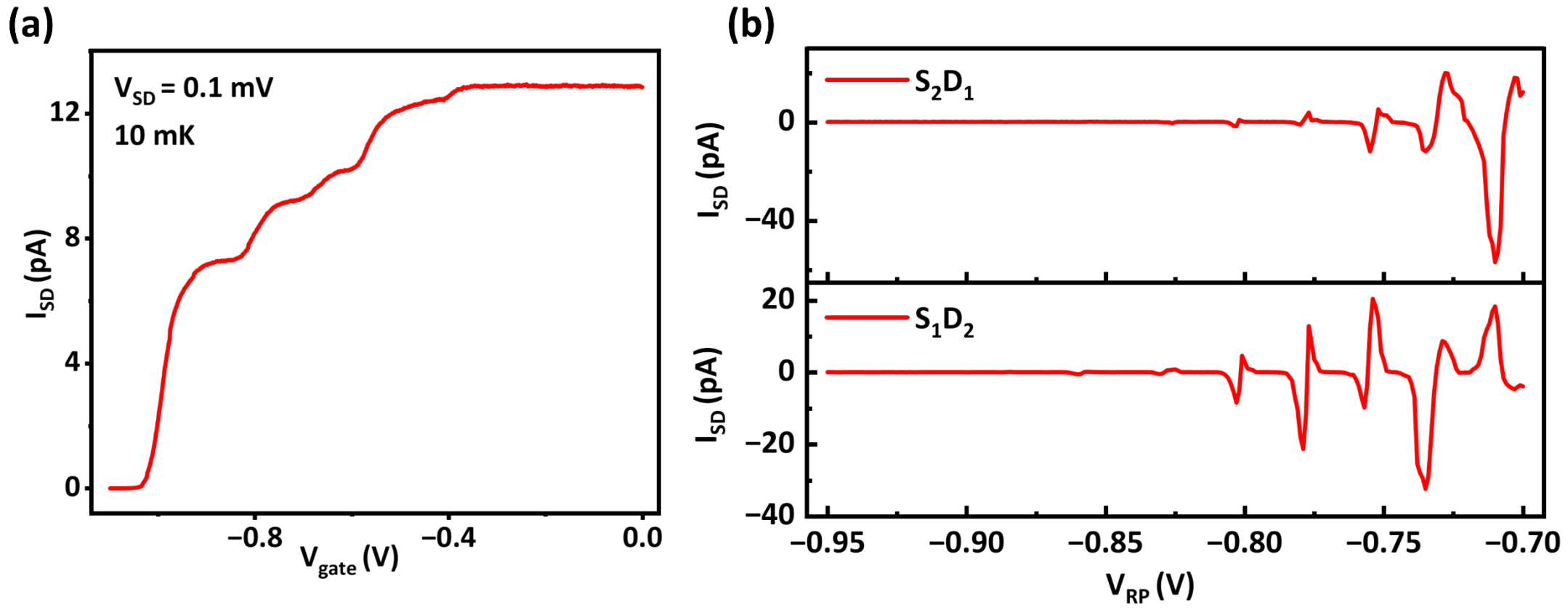

**Figure S1.** (a) Transfer characteristics of the gate pair CB-MB with $V_{gate}$ = $V_{CB}$ = $V_{MB}$ at a cryostat temperature of 10 mK, showing a clear pinch-off of the channel. (b) Plunger gate sweeps in the same barrier gate configuration at $V_{SD}$ = 0, for the two different directions of source-drain bias. $S_iD_j$ denotes using the ith and jth ohmic contacts as source and drain respectively.

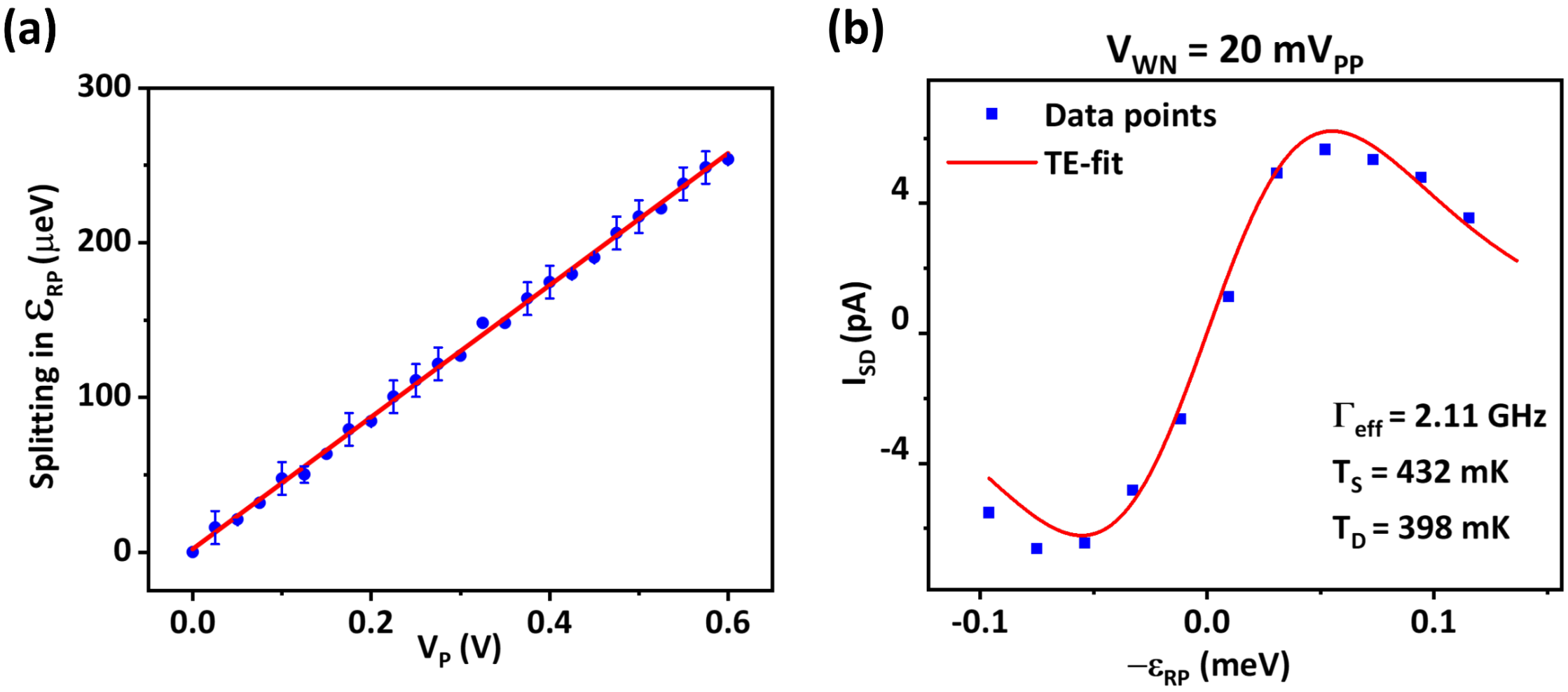

**Figure S2.** (a) Linear fit to the splitting of the bias-driven CBO peak at $\varepsilon_{RP}$ = 0 in Figure 4(a) with gate pulse amplitude $V_P$, giving the calibration of $V_P$ to energy $\varepsilon_P$ as: $\varepsilon_P$ (in meV) = 0.00196 + [0.42602 * $V_P$ (in V)]. (b) TE-fit to the $I_{SD} - \varepsilon_{RP}$ sweep at $\varepsilon_P$ = 0 for the TE-feature at $\varepsilon_{RP}$ = 0 in Figure 4(b), giving the electron temperatures.

[*] Corresponding author, madhu@iisertvm.ac.in

**S3:** Equation for $I_{sim}$

We derive the QD current when a square gate pulse of amplitude $V_P$ (of energy $\varepsilon_P$) and a D.C. gate voltage $V_{RP}$ (of energy $\varepsilon_{RP}$) are applied at the plunger gate RP, for the case where transport is thermally driven and only ground-state transitions are allowed.

- $I_{SD}$ through a single energy level in the presence of the square gate pulse:

With $V_P$ shifting a QD level of electrochemical potential $\mu_N$ to the new positions ($\mu_N - \varepsilon_P/2$) and ($\mu_N + \varepsilon_P/2$) for equal durations during one cycle, $I_{SD}$ has contributions from the two corresponding terms, and the total TE-current at $V_{SD} = 0$ is given by

$$\frac{I_{SD}\,(\mu_N,\ \varepsilon_{RP},\ \varepsilon_P)}{e\,.\,\Gamma_{eff}\,/\,2} = [f_S - f_D]\left(\varepsilon_{RP} + \mu_N + \frac{\varepsilon_P}{2}\right) + [f_S - f_D]\left(\varepsilon_{RP} + \mu_N - \frac{\varepsilon_P}{2}\right) \qquad \text{(S3-1)}$$

where the factor of ½ in the denominator of the left-hand side of the equation gives equal and reduced weightage to the two terms. The energy axis along the Fermi distribution at the leads is experimentally accessed by changing the QD electrochemical potential through $V_{RP}$ and $V_P$.

- Net QD current at a point in the $\varepsilon_{RP} - \varepsilon_P$ parameter space of the experimental data of Figure 4(b):

We consider the three successive charge transitions at $\varepsilon_{RP}$ = 0.4279, 0, and -0.5238 meV (shown in Figure 4(b)). The total TE-current $I_{sim}$ at a point in the $\varepsilon_{RP} - \varepsilon_P$ parameter space is calculated by summing over the individual TE-currents due to the three QD resonances, with a rough scaling factor to account for the difference in tunnel rates (and hence the $T_S$, $T_D$ values).

Table-I gives the scaling factor calculation at the three QD resonances. The average of the maximum amplitudes observed in positive ($I_{+,\,max}$) and negative ($I_{-,\,max}$) directions for each resonance is calculated at $V_{GP} = 0$, and normalised with respect to the value at $\varepsilon_{RP} = 0$ to get the scaling factor.

| $\varepsilon_{RP}$ (meV) | $I_{+,\,max}$ (pA) | $\lvert I_{-,\,max}\rvert$ (pA) | $(I_{+,\,max} + \lvert I_{-,\,max}\rvert)/2$ (pA) | Scaling factor relative to $\varepsilon_{RP} = 0$ |
|---|---|---|---|---|
| 0.4279 | 14 | 21 | 17.5 | 2.92 |
| 0 | 7 | 5 | 6 | 1 |
| -0.5238 | 3.4 | 6 | 4.7 | 0.78 |

Table-I. Scaling factor calculation to adjust for the difference in $\Gamma_{eff}$ between the three resonances shown in Figure 4(b).

The total TE-current $I_{sim}$ is thus given by

$$I_{sim}(\varepsilon_{RP},\ \varepsilon_P) = 2.92 * \frac{I_{SD}\,(0.4279,\ \varepsilon_{RP},\ \varepsilon_P)}{e\,.\,\Gamma_{eff}\,/\,2} + \frac{I_{SD}\,(0,\ \varepsilon_{RP},\ \varepsilon_P)}{e\,.\,\Gamma_{eff}\,/\,2} + 0.78 * \frac{I_{SD}\,(-0.5238,\ \varepsilon_{RP},\ \varepsilon_P)}{e\,.\,\Gamma_{eff}\,/\,2} \qquad \text{(S3-2)}$$

where $\frac{I_{SD}\,(\mu_N,\ \varepsilon_{RP},\ \varepsilon_P)}{e\,.\,\Gamma_{eff}\,/\,2}$ is as defined in Equation S3-1.

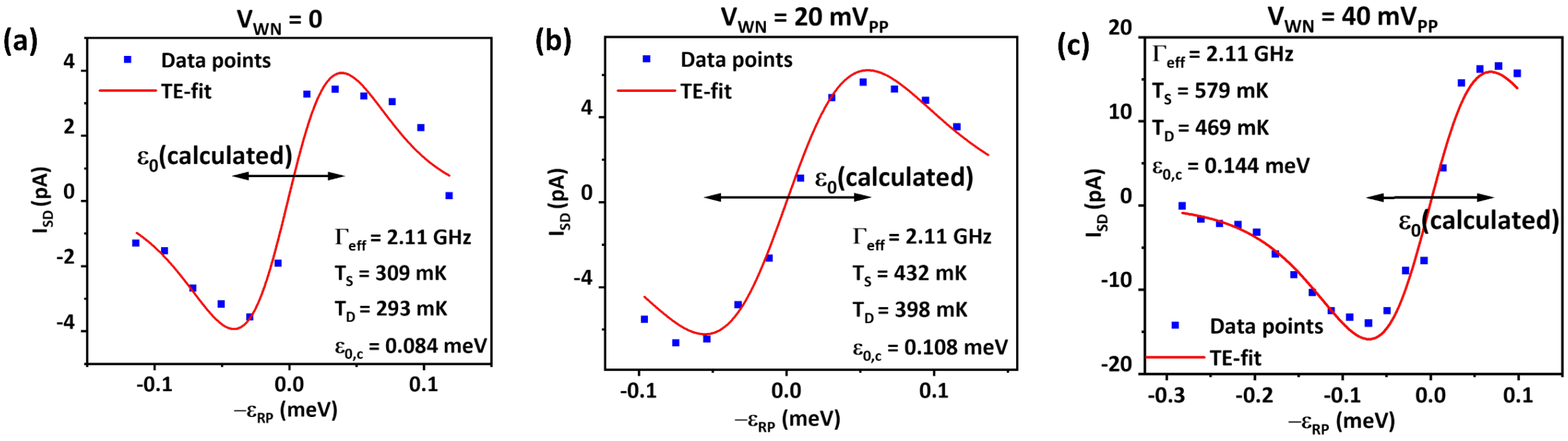

**Figure S4.** TE-fits to the $I_{SD} - \varepsilon_{RP}$ sweeps at $\varepsilon_P = 0$ for the TE-features (a) at $V_{WN} = 0$ of Figure 4(d), (b) $V_{WN} = 20\ mV_{PP}$ of Figure 4(e), and (c) $V_{WN} = 40\ mV_{PP}$ of Figure 4(f), giving the electron temperatures. The value of $\varepsilon_0$ calculated from the fitted curves is mentioned in each case.